# Transient stability of a power system with superconducting fault current limiters

**V. Sokolovsky, V. Meerovich**
Physics Department of the Ben-Gurion University of the Negev, Beer-Sheva, Israel
**I. Vajda**
Department of Electrical Machines and Drives, Budapest University of Technology and Economics, Budapest, Hungary

**Abstract.** The influence of a superconducting fault current limiter (FCL) on the transient stability of the synchronic operation of electric machines is analyzed for different locations of the inductive FCL in a network and for different parameters of the device. It is shown that the stability can be improved or degraded depending on the FCL impedance under a fault and the time of the recovery of the initial state of the limiter after a fault. Improving the transient stability with the inductive superconducting FCL is demonstrated in the experiments on the electrodynamic model of a power system. The expansion of the obtained results for other FCL designs is discussed.

## 1. Introduction

Application of fault current limiters (FCLs) based on high-temperature superconductors is one of the promising solutions of the fault current problem in power electric systems. Several power prototypes of the FCLs have been successfully tested [1-5]. The characteristics were achieved that could allow applying them in the distribution networks. At present there are developments directed to design superconducting FCL for application in high voltage lines [6]. The intense interest in these FCLs is explained by achievements of the technology of high-temperature superconductors and anticipated payoff from their application in power systems. The requirements to FCL parameters and to properties of superconductors are discussed in many publications (for example see [5, 7-11] and reverences noted in them). The requirements to FCLs can be separated according to their influence on electromagnetic processes in the system and influence on stability of parallel operation of electrical machines. It was shown that FCLs based on high-temperature superconductors meet to the first type requirements: they have low impedance under the normal operation regime of the protected circuit; under a fault their impedance fast increases limiting the first peak of a fault current and its steady-state value without appearance of dangerous overvoltages; FCLs quickly return into the initial low impedance state after the limitation of fault currents. Usually parameters and installation places of FCLs are chosen to ensure the required current limitation.

An FCL application has not to lead to worsening the static and transient stability of the power system [10]. An ideal limiter has zero impedance under the normal conditions of the circuit to be protected. In reality, it is enough that the voltage drop across an FCL under the normal conditions of a circuit is less than several percents (usually 5%) of the rated circuit voltage. In this case the FCL does not disturb the static stability of the power system operation and does not influence handling properties of the lines.

There are several studies devoted to analysis of the transient stability of a power system where superconducting FCLs are installed [10,12-19]. In these investigations the attention is given to two FCL designs, resistive and inductive, installed in one of the parallel transmission lines. It has been shown that the FCLs not only limit a fault current but also can increase the stability of the synchronic operation of the electric machines. However the carried out investigations leave open question: How does the influence of FCL on the system transient stability dependent on its installation place and parameters such as impedance in the limitation regime and recovery time into the initial low impedance state after limitation?

In this paper, we analyze in detail the influence of superconducting inductive FCLs on the transient stability of the power system. The FCL influence on the transient stability was analyzed using the area law. On the one hand this approach clearly shows how the installation place of a device and its parameters change the transient stability of power systems. On the other hand, this allows us to extend the obtained results for some other FCL designs (see Section 5). All the qualitative results were confirmed by the numerical simulation of transient processes in the power system. Some of the simulation results we published early [10,12,14]

## 2. Mathematical FCL model for transient stability analysis

The operation principle of most developed superconducting FCLs is based on the transition of active elements from the superconducting into the normal state under a fault condition in the protected circuit. Here we consider an inductive FCL, the secondary coil of which is closed by an active element [1,2,10]. Under the normal regime of the protected circuit an active superconducting element of the FCL is in the superconducting state, the device impedance $z_\sigma$ is low and the FCL does not influence on the nominal operation regime of the power system. The transition of the active element into the normal state under a fault causes a fast increase (during 0.003-0.01 s) of the device impedance up to a value $z_r$ needed for a limitation. This leads to the limitation of transient and steady-state fault currents. During the limitation regime the active element is in the normal state and it is heated. Therefore recovery of the superconducting state takes some time after removal of the current overload. If a FCL is inserted to a power line where a no-current pause is realized, the optimum thermal regime of the active element is a regime providing the recovery of the superconducting state during the pause. If a no-current pause cannot be realized, the recovery of the superconducting state is accompanied by alternate phase transitions into the superconducting state and back into the normal one [4,10]. Every phase transition is accompanied by an electromagnetic transient process in the protected circuit. To determine the influence of these transitions on the electromagnetic torque of the generator, we considered a single line with an inductive FCL simulated by the equivalent circuit of a two-coil transformer with a non-linear resistance in the secondary coil. This resistance modeled the superconducting active element with the account of the dependence of its resistance on a current and temperature [4, 10]. The calculation schema is based on a simultaneous numerical solution of the circuit equations with a non-linear resistance and heat equation describing the thermal state of the superconducting active element (details of the mathematic model see in [10]).

The electromagnetic torque influencing a rotor of the machine was taken proportional to a sum of products of the phase emfs and currents. Fig. 1 presents the calculation results of the generator torque after a one-phase fault clearing. The relative magnitude of the impedance of the FCL is 0.5. The result is compared with the simulation for a linear electrical reactor with constant impedance equaled to the FCL impedance when its active element is in the normal state. For chosen parameters, the active element is in the superconducting state with zero resistance a quarter of every period. Nevertheless the generator torque is practically the same as for a linear reactor with constant impedance. The difference between the torques calculated for the linear reactor and the FCL with a non-linear characteristic reduces more with decreasing the FCL impedance or by increasing the number of parallel lines. Thus, to analyze the transient stability, the inductive FCL can be described by a simplified mathematical model. In the framework of this model, under nominal operating conditions of the circuit, the FCL impedance is equal to a low value $z_\sigma$, and under a fault regime it jumps up to $z_r$ keeping unchanged until the superconducting state is reestablished.

### 3. Qualitative analysis

Evaluation of the FCL influence on the transient stability behavior is carried out in accordance with the equal area criterion. The transient stability is kept if the acceleration area is equal or less the deceleration area. The FCL influence on the transient stability is determined by the comparison of the ratio of these areas with and without an FCL.

3.1. *A block of generator-transformer connected by a single line to an infinite bus*
Let us consider FCLs installed in the line as shown in Fig. 2. The transmitted power from the generator to the system is given by well-known expression

$$P = EU \sin(\delta)/X, \qquad (1)$$

where $E$ is the generator emf, $U$ is the bus voltage, $\delta$ is the angle between the vectors $U$ and $E$, $X$ is the total interconnective reactance. Here we assume that active part of the impedances of all the devices in the circuit can be neglected and also $z_\sigma = x_\sigma$; $z_r = x_r$. (Under nominal operating conditions of the circuit to be protected the active part of the FCL impedance equals to a resistance of the primary coil and is much less than the inductive part. Under a fault, a resistance of the active element is much higher than the inductive part of the FCL impedance and can be neglected at determination of the total interconnective reactance).

Analysis of the transient stability is usually performed for two cases: a one-phase fault as the most frequent case in a high-voltage line and a three-phase fault as the worst case. In the case of a three-phase fault in the point **A**, Fig. 2, $P = 0$ and the generator rotor accelerates, the angle $\delta$ increases from $\delta_0$ at the nominal regime to $\delta_2$ (Fig. 3). The angle $\delta_1$ corresponds to opening the circuit breaker $S_1$ and the angle $\delta_2$ - to reclosing the breaker. If the FCLs come back into the initial state during a no-current pause, the squares of both the acceleration **AA** and damping **DA** areas do not change in comparison with these areas, when FCLs are not installed in the circuit (Fig. 3a). If the FCLs come back into the initial state later reclosing the breaker $S_1$, the square of the acceleration area **AA** is the same but the damping area square **DA** is less (Fig. 3b). Due to increasing the circuit impedance up to $X + x_r$ between $\delta_2$ till $\delta_3$, where the initial state of the FCL recovers, power $P$ reduces in comparison with the case without FCLs. The transient stability decreases. The same situation is observed at a three-phase fault in point **B** (Fig. 2). Power $P = 0$ until circuit breaker $S_2$ opens the circuit and the power is less with FCLs until they return to the initial state. Note, that the description is valid and the transient stability decreases where FCLs are installed between a transformer and a bus or between a generator and a transformer.

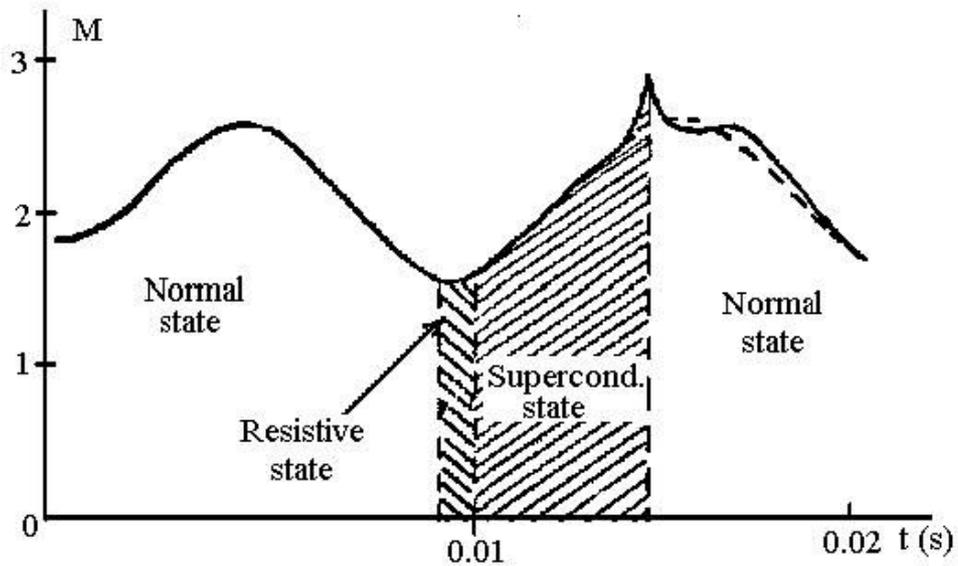

Fig. 1. Temporary dependence of torque. Calculation with account of the superconducting – normal state transitions (solid curve) and according the simplified model (dashed curve).

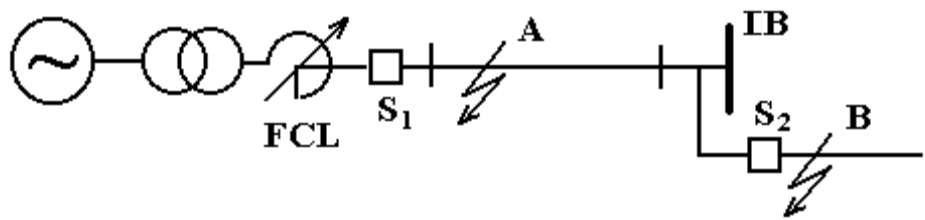

Fig. 2. Simplest circuit for the transient stability analysis: a generator-transformer block connected to an infinite bus with an FCL installed in the line. $S_1$ and $S_2$ –circuit breakers.

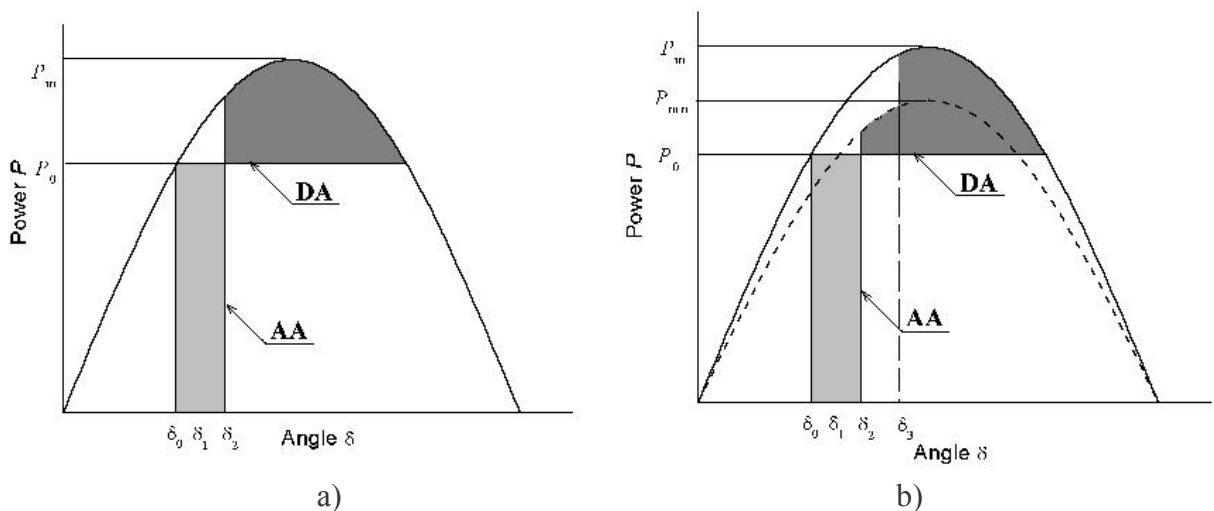

Fig. 3. Power diagrams for analysis of the transient stability at a three-phase fault in the point **A** of the circuit shown in Fig. 2: (a)- without FCLs; (b)- with FCLs.

At a one-phase fault the currents in healthy phases increase also [20]. This can initiate the activation of the FCLs installed in these phases and, following, the impedances increase in every phase. The analyses will be based on the equivalent circuit, which can be presented in the form shown in Fig. 4. Based on the positive sequence equivalent criterion, while a short circuit occurs, the use of the symmetrical components allows any type of faults to be represented in the positive sequence network by a fault shunt $x_f$ connected between the point of the fault and the neutral. The value of the shunt depends on the type of fault [17,21]. For a one-phase fault event in point **A** (Fig. 2), when all the FCLs activates the total interconnective reactance $X_1$ is:

$$X_1 = x_t + x_r + x_l + (x_t + x_r) x_l / x_f, \qquad (2)$$

where $x_f = (x_t + x_r)//x_l + (x_{t0} + x_r)//x_{l0}$ ; $x_t$, $x_l$ are the positive sequence impedances of the generator-transformer block and line, respectively; the symbol "//" notes the parallel connection of the impedances; "0" marks the impedances of the zero sequence; the FCL impedance of the zero sequence equals to the positive sequence of the FCL.

Our estimations show that the circuit impedance with FCLs $X_1$ ($x_r > 0$) is larger than without limiters $X_1$ ($x_r = 0$) for real values of the impedances of the circuit and FCLs. The acceleration area increases due to the decrease of power $P$ during the fault and due to the increase of the angle corresponding to opening the circuit breaker (accelerating torque grows). At the same time, the damping area reduces since some time is needed to recover the FCLs installed in healthy phases into the initial state with the low impedance $x_\sigma$.

Analogous reasoning shows that the transient stability reduces at a one-phase fault in point **B**.

If only a single FCL in a faulty phase is activated at a one-phase fault, there are two competitive influences of FCL on the transient stability: during a fault a FCL decreases a interconnective reactance and, following, the square of the acceleration area **AA** is decreased; but if the FCL does not return into the initial low impedance state during a no-current pause, the interconnective reactance increases and the damping area square **DA** is decreased. Our numerical simulation shows that the influence of the FCL is determined by the following value

$tx = x_r(t_r - t_p)$

where $t_r$ is the time needed for the FCL to return to the low impedance; $t_p$ is the duration of a no-current pause.

The stability is improved if this value is below a particular critical value; in the opposite case the transient stability decreases. The critical value is determined by the parameters of the protected circuit and the inertial moment of the generators.

## 3.2. Power station containing several generator-transformer blocks

Now let us consider cases of the FCL installation in a power station where several blocks are connected to a common bus. The station is connected to a power system via several lines (Fig. 5a). In the first case marked as **1** in Fig. 5a, FCLs are installed between the transformers and the station bus. A three-phase fault in one of the lines near the bus (point **A**) activates all the FCLs but $P = 0$ during the fault, as in the discussed above case with a single line. As distinct from the single line case, during a no-current pause, when the circuit breaker in a faulty line is open, the impedance $X$ is higher in comparison with the impedance of the system without limiters and, hence, the power $P$ is less until the FCLs return into the initial state. Thus the installation of the FCLs decreases the transient stability of the power system at a three-phase fault in the line. For the case of a one-phase fault, we have to take into account that both the healthy phases of the line and other lines shunt the faulty phase. This changes the critical value of $tx$. The qualitative picture of the FCL influence on the stability is similar to a one-phase fault in a single line with FCLs.

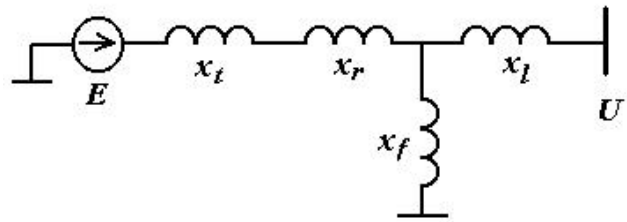

Fig. 4. Equivalent circuit of the configuration presented in Fig. 2.

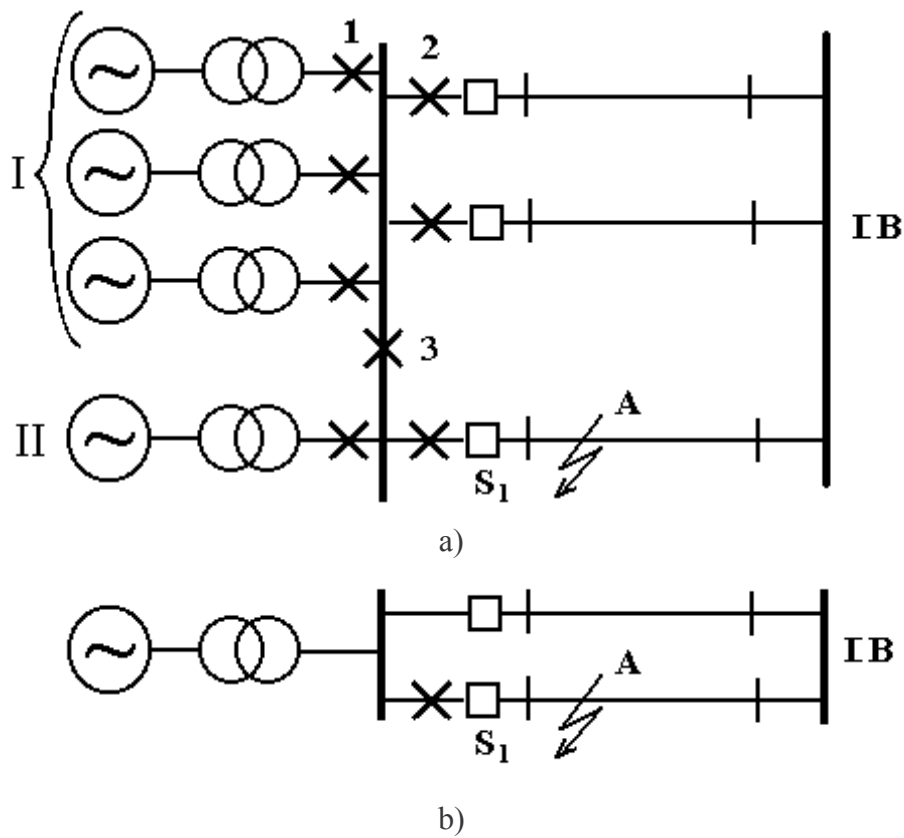

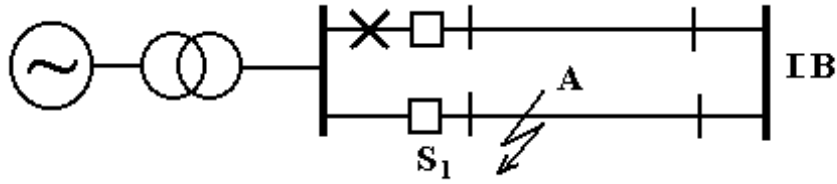

c)

Fig. 5. Circuit with several generators connected to a common infinite bus with several lines. $S_1$ and $S_2$ –circuit breakers. **X** marks possible places of FCL installation. Simplified schemes for generator group I (b) and group II (c). Here **X** marks the activated FCLs.

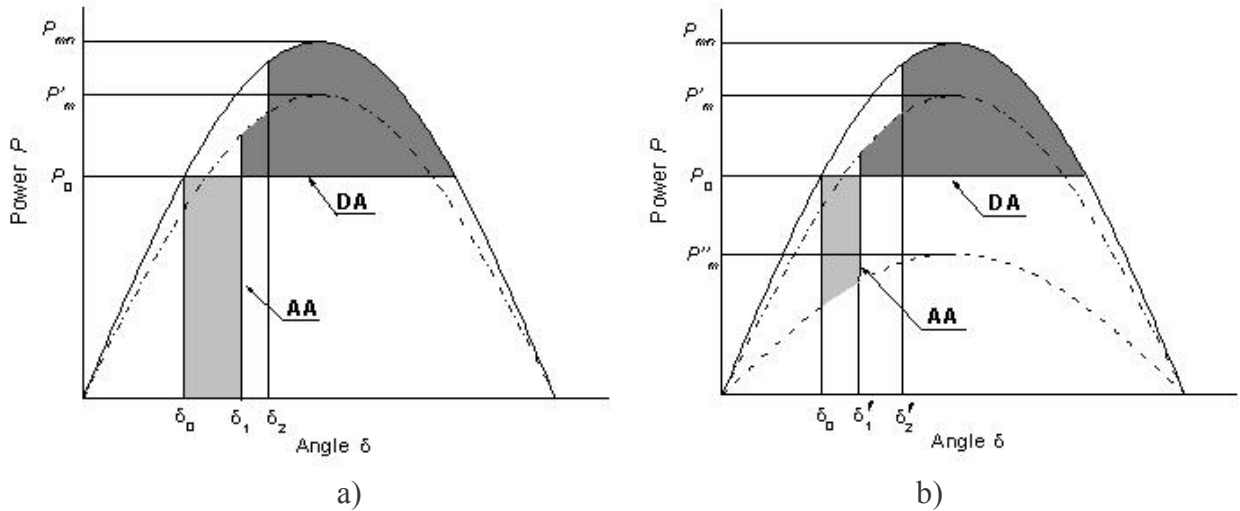

a)            b)

Fig. 6. Power diagrams for analysis of transient stability at a three-phase fault in the point **A** of the circuit shown in Fig. 4: (a)- without FCLs; (b)- with FCLs.

The second possible case of the FCL installation is in the lines connecting the station and the system (the FCL location marked by **2** in Fig. 5). Let us consider a three-phase fault in one of the lines (point **A**). The power diagram for the analysis of the transient stability without FCLs is presented in Fig. 6a. A fault occurs at angle $\delta_0$ corresponding to the nominal regime with $P = P_m \sin(\delta_0)$. The power $P = 0$ until the line circuit breaker $S_1$ opens the line at the angle $\delta_1$. During the no-current pause in the faulty line the power is $P = P'_m \sin(\delta)$ with the impedance $X'$ determined by the healthy lines. Only after reclosing the line breaker at the angle $\delta_2$ $P = P_m \sin(\delta)$.

The power diagram with FCLs is presented in Fig. 6b. As differentiated from the first case, only the FCLs in the faulty line come into action. In this case power $P$ does not fall to zero at a three-phase fault at point **A**. During the fault $P = P''_m \sin(\delta)$ with the impedance $X''$ that can be calculated using the equivalent schema similar to one shown in Fig. 4. However this time the device impedance $x_r$ is absent in the left hand part of the schema, $x_f = x_r$, and in the right hand part $x_l$ is determined as the impedance of the parallel connected healthy lines. As one can see from comparison of Figs. 6a and 6b, the acceleration area decreases due to decreasing accelerating torque. It leads to reducing angles $\delta_1$ and $\delta_2$ corresponding to opening and reclosing the line by circuit-breaker $S_1$. If the FCLs return into the initial state with the low impedance during the no-current pause, the damping area square increases and the transient stability is improved. If the FCLs do not return into the initial state at the moment of reclosing the line, it increases the impedance *X* and decreases the damping area. In this case the transient stability can be decreased alike it was discussed for cases of a one-phase fault in Section 3.1 and above three-phase fault when FCLs are placed between the generators and bus. Note, that it was assumed that FCLs in the healthy lines did not activate due to fault currents are flowing from the infinite bus. Otherwise, there are two competitive effects from FCLs: a decrease of the total interconnective reactance *X* and of the acceleration area **AA** during a fault and on the other hand, an increase of the reactance till the FCLs return into the initial low impedance state after a fault. This leads to decrease of the damping area square **DA** and, following, can make worse the transient system stability. Our numerical simulation shows that the influence of the FCL is determined by the FCL parameters (FCL impedance in limitation regime and recovery time of the initial state) and parameters of the protected system (the number of the generator-transformer blocks and lines, their impedances, operation speed of the circuit breakers).

The third possible case of the FCL installation is in the connecting tie of the buses the (the FCL location marked by **3** in Fig. 5a). For qualitative analyses let us assume that power and inertia of the generator group I is much higher than of the group II. In this case the analysis of the transient stability can be simplified and be carried out for each group separately. The simplified schemes for the transformer groups I and II are presented in Figs 5b and 5c, accordingly. Analyses of the transient stability for the group I (Fig. 5a) is completely coincided with the discussed above case where FCLs are installed in every line and the FCLs are activated only in a faulty line.
For the generator group II the situation is dramatically different: the FCLs are activated in the healthy line (Fig. 5c) and are not activated in the faulty line. This leads to increase of the total interconnective reactance during any fault as well as during a no-current pause and also after a reclosing the breaker **S₁** till the FCLs return into the initial low impedance state. As result, the damping area square **DA** and stability are decreased.

### 4. Experimental investigation of transient stability

The qualitative analysis shows that, in *definite* cases, the FCL installation can increase of the transient stability. These results were confirmed in the experiments on the electrodynamic model of a power system containing a synchronic machine of 3.4 kW, two three-phase 5.4 kVA transformers, a set of elements modeling a power line, controlled switches for modeling a fault event, and circuit breakers. The nominal line voltage and current were 500 V and about 1 A, respectively. A model of the inductive superconducting FCL was installed in one phase of the line near a block of the machine-transformer. The model was built on base of an open magnetic system [10, 12]. The impedance of the FCL increased from $x_\sigma$ = 5.6 Ohm up to $x_r$ = 26 Ohm during about 0.003 s when an instantaneous current in the circuit exceeded the activation value of about 3.2 A. The model limited a current from 23 A down 11 A at a one-phase fault and recovered the initial state during a no-current pause. Fig. 7 presents in sequence the oscilloscope traces of a one-phase fault without FCL, with a linear reactor having the impedance of 5.6 Ohm and with the FCL model. One can see that only the FCL keeps the transient stability of the system.

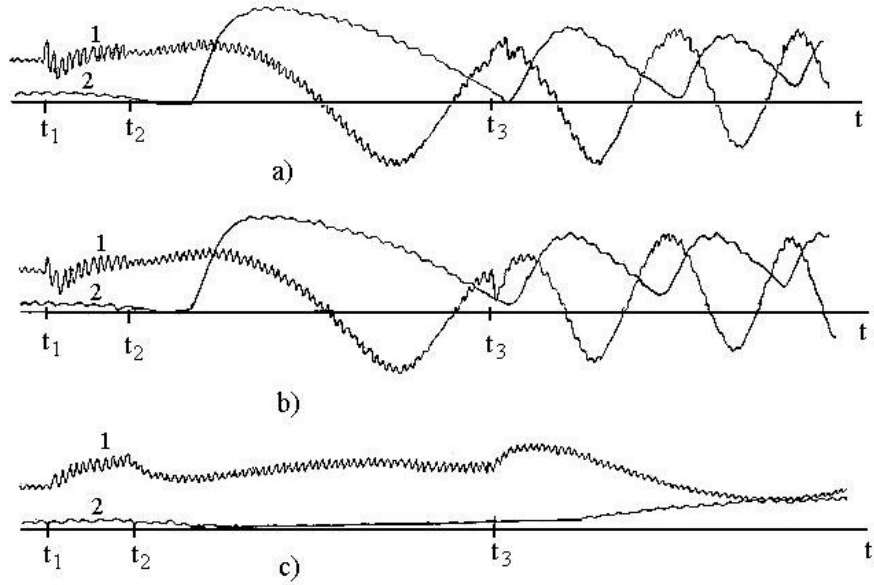

Fig. 7. Oscilloscope traces of a one-phase fault in the experimental circuit: (a) - without FCL; (b) - with linear reactor $x = 5.6\ \Omega$ ; (c) – with FCL. 1- power; 2- angle; $t_1$ – instant of fault; $t_2$ – time of opening faulty phase; $t_3$ – time of reclosing circuit-breaker.

## 5. Discussion

The obtained qualitative results have been confirmed by lot of numerical simulations based on the complete mathematic model as well as the simplified model. The complete model is based on a numerical simultaneous solution of the equations describing the relative motion of the generator rotors, electromagnetic processes in three-phase circuit and thermal state of the active HTS elements. Inductive FCLs are simulated by the equivalent circuit of a two-coil transformer the secondary coil of which is short-closed by a non-linear active element [4, 10]. In the simplified model the FCLs are simulated as a non-linear inductive element. The impedance of the element is equal to a low value $x_\sigma$ in the nominal regime and under a fault regime the impedance jumps up to $x_r$ (Section 2). The simulation based on these models gives close results and some of them are published by us wherein [10,14]. These results are similar to the simulation results obtained by other investigators [13,15]. For cases where our qualitative analyses predict an increase of the transient system stability the numerical simulations show that there are regimes when the transient stability is kept with FCLs and is disturbed without FCLs at the same time. It means that there are regimes where a relative angel infinitely increases without FCLs and returns to the initial value after damped oscillations with FCLs [10,13-15]. For cases where FCLs decrease the transient stability the reversed situation was obtained with help of numerical simulation. Some qualitative and simulation results are already confirmed experimentally (Section 4 and [16]).

Note, that application of the traditional methods of current limitation, such as using transformers with increased impedance, linear reactors, stationary and forward partition of power systems, leads to increase of the total interconnective reactance in fault regimes as well as in nominal regimes. Using the forward partition of the system causes also raising of duration of a fault. Therefore, in contrast to superconducting FCLs these methods decrease the transient stability of power system.

The application of inductive FCLs can increase the transient stability only in *definite* cases. The FCL influence depends strongly on the parameters and place of the installation. Using the obtained results, the following recommendations can be formulated to increase the transient stability:
- install FCLs in the lines;
- chose the activation current of FCLs higher than the increased current in the healthy lines and phases at one-phase fault;
- provide the recovery of the superconducting state of the active element and return of the FCLs to the initial state with a low impedance during a no-current pause;
- chose the activation current of FCLs higher than a current of "long-distance" fault event (current of a fault in point **B**, Fig. 2).

The FCLs meeting these requirements increase the transient stability of the power system at a fault in the lines and does not change the stability at a fault inside the system. Applicability of the obtained results and requirements can be extended for other designs of fault current limiters including non-superconducting. From the point of view of the transient stability these devices have to be modeled by a nonlinear reactor with a low reactance under the nominal regime of the protected circuit. Under a fault event the impedance fast increases (jumps) up to a high value keeping unchanged during the fault and some time after. For example, the following designs are met to the requirements: superconducting resistive FCLs shunted by a coil [10], HTS fault current controller [23,24], superconducting and traditional FCLs with saturable magnetic core [25,26]. Note the first of them, resistive FCL, as like an inductive one, requires in some time $t_r$ to return into the initial time after interruption of a fault current. Hence, all the results obtained for an inductive FCL are applied to a resistive FCL where an active element is shunted by a coil. The existence of the time $t_r$ leads to application of FCLs can decrease the transient stability, for example, in the case of a three-phase fault in a line of the system where FCLs are installed between a transformers and bus. The FCLs with a saturable magnetic core and fault current controllers can return into the initial low impedance state immediately after the interruption of a fault current, without any delay. Absence of delay makes better the transient stability of power system with FCLs in comparison with the case of an inductive FCL. For example, installation of FCLs between a transformers and bus does not decrease the transient stability at a three-phase fault in a line. However, the requirements to a value of the activation current remain the same.

The obtained results can be applied for analyses of the transient stability of power rotating load. The stability of the parallel operation of synchronous motors can be also analyzed with help of the area law. In this case the power diagrams are symmetrical to the ones discussed above. The numerical simulation [10, 12] has shown that inductive HTS FCLs installed in lines increase the stability of the parallel operation of synchronous motors.

**References**

[1] W. Paul, M. Lakner, J. Rhyner, P. Unternahrer, Th. Baumann, M. Chen, L. Windenhorn, and A. Guerig, "Test of a 1.2 MVA high-Tc superconducting fault current limiter," in *Inst. Phys. Conf. Ser.*, vol. 158, pp. 1173-1178, 1997.

[2] V. Meerovich, V. Sokolovsky, G. Jung, and S. Goren, "High-Tc superconducting inductive current limiter for 1 kV/25A performance," *IEEE Trans. Appl. Superconduct.*, vol. 5, pp. 1044-1048, June 1995.

[3] B. Gromoll, G. Ries, W. Schmidt, H.-P. Kraemer, B. Seebacher, B. Utz, R. Nies, H.-W. Neumueller, E. Baltzer, S. Fisher, and B. Heismann, "Resistive fault current limiters with YBCO films-100 kVA functional model," *IEEE Trans. Appl. Superconduct.*, vol. 9, pp. 656-659, June 1999.

[4] R. Witzmann, W. Schmidt, and R.-R. Volkmar, "Resisteve HTSL-strombegrenzer", presented at *Internationaler ETG-Kongress 2001*, 23-24 Oct., 2001, Nuremberg, Germany.

[5] Roadmap for Europe (2/06/01), in *Proc. Fault Current Limiter Working Group Workshop, SCENET,* Grenoble, CNRS, pp. 5-28, May 2001.

[6] T. Yazawa, H. Koyama, K. Tasaki, T. Kuriyama, S. Nomura, T. Ohkuma, N. Hobara, Y. Takahashi, and K. Inoue, "66 kV-Class high-Tc superconducting fault current limiter magnet model coil experiment", *IEEE Trans. Appl. Superconduct.*, vol. 13, pp. 2040-2043, June 2003.

[7] R. F. Giese, "Fault current limiter –A second look", Argonne Nat. Lab.: Rep. for International Energy Agency (IEA), March 1995.

[8] L. Salasoo, A. F. Imece, R. W. Delmerico, and R. D. Wyatt, "Comparison of superconducting fault limiter concepts in electric utility applications", *IEEE Trans. Appl. Superconduct.*, vol. 5, pp. 1079-1082, June 1995.

[9] E. Leung, "Surge protection for power grids", *IEEE Spectrum,* vol. 34, pp. 26-30, July 1997.

[10] V. Sokolovsky, V. Meerovich, I. Vajda, and V. Beilin, "Superconducting FCL: Design and Application", *IEEE Trans. Appl. Superconduct.*, (in print 2004).

[11] K. Hongesombut, Y. Mitani, and K. Tsuji, "Optimal location assignment and desing of superconducting fault current limiters applied to loop power systems", *IEEE Trans. Appl. Superconduct.*, vol. 13, pp. 1828-1831, June 2003.

[12] V. Sokolovsky, V. Meerovich, and I. Vajda, "Superconducting fault current limiter: state of development and possible applications," in book *"Advanced Studies on Superconducting Engineering"*, Eds. I. Vajda, L. Farkas, Hungary, 2001, pp. 253-271.

[13] S. Lee, Ch. Lee, T. K. Ko, and O. Hyun, "Stability analysis of power system with superconducting fault current limiter installed," *IEEE Trans. Appl. Supercond.,* vol. 11, pp. 2098-2101, 2001.

[14] I. Vajda, A. Gyore, A. Szalay, V. Sokolovsky, W. Gawalek, "Improved Desing and System Approach of a Three Phase Inductive HTS Fault Current Limiter for 12 kVA Synchronous Generator", *IEEE Trans. Appl. Superconduct.*, vol. 13, pp. 2000-2003, June 2003.

[15] M. Sjöström, R. Cherkaoui, and B. Dutoit, "Enhancement of power system transient stability using superconducting fault current limiters", *IEEE Trans. Appl. Superconduct.*, vol. 9, pp. 1328-1330, June 1999.

[16] M. Tsuda, Y. Mitani, K. Tsuji, and K. Kakihana, "Application of resistor based superconducting fault current limiter to enhancement of power system transient stability", *IEEE Trans. Appl. Superconduct.*, vol. 11, pp. 2122-2125, March 2001.

[17] L. Ye, L.Z. Lin, and K.-P. Juengst, "Application studies of superconducting fault current limiters in electric power systems", *IEEE Trans. Appl. Superconduct.*, vol. 12, pp. 900-903, March 2002.

[18] H. Hatta, S. Muroya, T. Nitta, Y. Shirai, and M. Taguchi, "Experimental study on limiting operation of superconducting fault current limiter in double circuit transmission

line model system", *IEEE Trans. Appl. Superconduct.*, vol. 12, pp. 812-815, March 2002.

[19] Y. Shirai, M. Taguchi, M. Shiotsu, H. Hatta, and T. Nitta, "Simulation study on operating characteristics of superconducting fault current limiter in one-machine infinite bus power system", *IEEE Trans. Appl. Superconduct.*, vol. 13, pp. 1822-1827, June 2003.

[20] I. Vajda, S.Semperger, T Porjesz, A. Szalay, V. Meerovich, V. Sokolovsky, W. Gawalek, "Thre Phase Inductive HTS Fault Current Limiter for the Protection of a 12 kVA Synchronous Generator", *IEEE Trans. Appl. Superconduct.*, vol. 11, pp. 2515-2518, March 2001.

[22] S. B. Losev and A. B. Chernin, "Calculation of electrical values at nonsymmetrical regime of electrical systems", Energoatomizdat, Moscow, 1983.

[23] M. Parizh and E. Leung, "Power quality, micro superconducting magnetic energy storage systems, and fault current limiters", in *Applications of Superconductivity*, Ed. H. Weinstock, NATO ASI Series, Series E: Applied Sciences, vol. 365, pp. 415-456, Kluwer Academic Publishers, 2000.

[24] J. A. Waynert, H. J. Boenig, C. H. Mielke, J. O. Willis, and B. L. Burley, "Restoration and testing of an HTS fault current controller", *IEEE Trans. Appl. Superconduct.*, vol. 13, pp. 1984-1987, June 2003.

[25] B. P. Raju, K. C. Parton, and T. C. Bartram, "Fault current limiting reactor with superconducting DC bias winding," *CIGRE* 1982, pp. 1-9, September, no. 23-03.

[26] J. X. Jin, S. X. Dou, H. K. Liu, C. Grantham, Z. J. Zeng, Z. Y. Liu, T. R. Blackburn, X. Y. Li, H. L. Liu, and J. Y. Liu, "Electrical application of high Tc superconducting saturable magnetic core fault current limiter," *IEEE Trans. Appl. Superconduct.*, vol. 7, pp. 1009-1012, June 1997.